\documentstyle[epsfig]{aipproc}

\begin{document}

\newcommand\eq[1]{Eq.~(\ref{#1})}
\newcommand\eqs[2]{Eqs.~(\ref{#1}) and (\ref{#2})}
\newcommand\eqss[3]{Eqs.~(\ref{#1}), (\ref{#2}) and (\ref{#3})}
\newcommand\eqsss[4]{Eqs.~(\ref{#1}), (\ref{#2}), (\ref{#3})
and (\ref{#4})}
\newcommand\eqssss[5]{Eqs.~(\ref{#1}), (\ref{#2}), (\ref{#3}),
(\ref{#4}) and (\ref{#5})}
\newcommand\eqst[2]{Eqs.~(\ref{#1})--(\ref{#2})}

\newcommand\ee{\end{equation}}
\newcommand\be{\begin{equation}}
\newcommand\eea{\end{eqnarray}}
\newcommand\bea{\begin{eqnarray}}
\renewcommand{\topfraction}{0.99}
\title{Superhorizon perturbations and preheating}
\author{Karim A.~Malik$^1$, David Wands$^1$, David H.~Lyth,$^2$ and
Andrew R.~Liddle$^{3}$}
\address{$^1$School of Computer Science and Mathematics, Mercantile House,
University of Portsmouth, Portsmouth PO1 2EG, U.~K.}
\address{$^2$Department of  Physics, University of Lancaster,
Lancaster LA1 4YB, U.~K.}
\address{$^3$Astronomy Centre, University of Sussex, 
Brighton BN1 9QJ, U.~K.}
\maketitle
\begin{abstract}
We discuss the evolution of linear perturbations about a
Friedmann--Robertson--Walker background metric, using only the local
conservation of energy--momentum. We show that on sufficiently large
scales the curvature perturbation on spatial hypersurfaces of
uniform-density is constant when the non-adiabatic pressure
perturbation is negligible. 
We clarify the conditions under which super-horizon curvature 
perturbations may vary, using preheating as an example.
\end{abstract}

Structure in the Universe is generally supposed to originate from the
quantum fluctuation of the inflaton field.  As each scale leaves the
horizon during inflation, the fluctuation freezes in, to become a
perturbation of the classical field. The resulting cosmological
inhomogeneity is commonly characterized by the intrinsic 
curvature of spatial hypersurfaces defined with respect to the matter.
This metric perturbation is a crucial quantity, because at approach of
horizon re-entry after inflation it determines the adiabatic
perturbations of the various components of the cosmic fluid, which
seem to give a good account of large-scale structure~\cite{LL}.

To compare the inflationary prediction for the curvature perturbation
with observation, we need to know its evolution outside the horizon,
through the end of inflation, until re-entry on each cosmologically
relevant scale.  The standard assumption is that the curvature
perturbation is practically constant. This has recently been called
into question in the context of preheating models \cite{KLS97} at the
end of inflation where non-inflaton perturbations can be resonantly
amplified \cite{Betal,LLMW}. 
We investigate the circumstances under which the curvature perturbation
may vary.

Using only the local conservation of energy--momentum, we show that
the rate of change of the curvature perturbation on uniform-density
hypersurfaces\footnote{The ``conserved quantity'' $\zeta$ was
originally defined in Bardeen, Steinhardt and Turner \cite{BST}, but
constructed from perturbations defined in the uniform Hubble-constant
gauge.}, $\zeta$, on large scales is due to the non-adiabatic part of
the pressure perturbation.  This result is independent of the form of
the gravitational field equations, demonstrating for the first time
that the curvature perturbation remains constant on large scales for
purely adiabatic perturbations in {\em any} relativistic theory of
gravity where the energy--momentum tensor is covariantly conserved.
In contrast with the usual approach to cosmological perturbation
theory, we shall not invoke any gravitational field equations. 
General coordinate invariance implies the energy-momentum conservation
law $T^{\mu}_{~\nu;\mu}=0$, without invoking the Einstein field
equations.

The pressure perturbation must be adiabatic if there is a definite
equation of state for the pressure as a function of density, which is
the case during both radiation domination and matter domination.  On
the other hand, a change in $\zeta$ on super-horizon scales will occur
during the transition from matter to radiation domination if there is
an isocurvature matter density perturbation~\cite{ks87,David+Tony}.

A simple intuitive understanding of how the curvature perturbation on
large scales changes, due to the different integrated expansion in
locally homogeneous but causally-disconnected regions of the universe,
can be obtained within the `separate universes' picture which we 
described in \cite{WMLL}. 

\section{Linear scalar perturbations}
\label{scalpert}

The line element allowing arbitrary linear scalar
perturbations of a Friedmann--Robertson--Walker (FRW) background can
be written~\cite{lifshitz,Bardeen,KS,MFB}
\be
\label{ds}
ds^2 = -(1+2A) dt^2+2a^2(t) \nabla_i B\, dx^i dt 
+ a^2(t)
\left[(1-2\psi)\gamma_{ij}+2\nabla_i\nabla_j E \right] dx^idx^j \, .
\ee
The unperturbed spatial metric for a space of constant curvature
$\kappa$ is given by $\gamma_{ij}$ and covariant
derivatives with respect to this metric are denoted by $\nabla_i$.

The curvature perturbation on fixed-$t$ hypersurfaces, $\psi$, 
is a gauge-dependent quantity and under an
arbitrary linear coordinate transformation, $t\to t+\delta t$, it
transforms as
$\psi \to \psi + H \delta t$. 
On uniform-density hypersurfaces the curvature perturbation 
can be written as\footnote{The sign of $\zeta$ is chosen here to
coincide with Refs.~\cite{BST}.}
\begin{equation}
\label{defzeta}
-\zeta = \psi +H\frac{\delta\rho}{\dot\rho}\, .
\end{equation}

The curvature perturbation on uniform-density hypersurfaces, $\zeta$,
is often chosen as a convenient gauge-invariant definition of the
scalar metric perturbation on large scales.
These hypersurfaces become ill-defined if the density is not strictly
decreasing, as can occur in a scalar field dominated universe when the
kinetic energy of the scalar field vanishes. In this case one can 
instead work in terms of the density perturbation on uniform-curvature
hypersurfaces, $\delta\rho_{\psi}=\delta\rho+\dot\rho\psi/H$, 
which remains finite.

The pressure perturbation (in any gauge) can be split
into adiabatic and entropic (non-adiabatic) parts,
by writing 
\begin{equation}
\delta p = c_{{\rm s}}^2 \delta\rho + \dot{p} \Gamma \, ,
\end{equation}
where $c_s^2\equiv \dot p/\dot \rho$. 
The non-adiabatic part is $\delta p_{\rm nad}\equiv \dot p \Gamma$,
and $\Gamma\equiv\delta p /\dot{p} - \delta\rho / \dot{\rho}$.
The entropy perturbation $\Gamma$, defined in this way, is
gauge-invariant, and represents the displacement between hypersurfaces
of uniform pressure and uniform density.

\section{Evolution of the curvature perturbation}
\label{curvpertsect}

The energy conservation equation $n^{\nu}T^\mu_{~\nu;\mu}=0$
for first-order density perturbations gives
\begin{equation}
\label{continuity}
\dot{\delta\rho} = -3H(\delta\rho + \delta p)
 + (\rho+p) \left[ 3\dot\psi - \nabla^2\left(v+\dot E\right) \right]
\, ,
\end{equation}
where $\nabla^i v$ is the perturbed 3-velocity of the fluid.
In the uniform-density gauge, where $\delta\rho=0$ and $\psi=-\zeta$,
the energy conservation equation~(\ref{continuity}) immediately gives
\begin{equation}
\label{dotzeta}
\dot\zeta = - {H\over \rho+p} \delta p_{\rm nad}
 - {1\over3} \nabla^2 \left(v+\dot E\right) \,.
\end{equation}
We emphasize that we have derived this result without invoking any
gravitational field equations, although related results have been
obtained in particular non-Einstein gravity
theories~\cite{earlierHwang}.
We see that $\zeta$ is constant if (i) there is no non-adiabatic
pressure perturbation, and (ii) the divergence of the 3-momentum on
zero-shear hypersurfaces, $\nabla^2(v+\dot E)$, is
negligible.

On sufficiently large scales, gradient terms can be neglected
and~\cite{GBW,David+Tony}
\begin{equation}
\label{dotzeta2}
\dot\zeta = - {H\over \rho+p} \delta p_{\rm nad} \,,
\label{keyresult}
\end{equation}
which implies that $\zeta$ is constant if 
the pressure perturbation is adiabatic. It has been argued 
\cite{David+Tony} that the divergence is likely to be negligible
on all super-horizon scales, and in  the following we shall make
that assumption.

\section{Preheating}

The chaotic inflation model studied originally in \cite{KLS97} 
and later on by \cite{Betal,LLMW} has the scalar field potential
$V(\phi,\chi) = \frac{1}{2} \, m^2\phi^2 +  \frac{1}{2} \, 
g^2\phi^2\chi^2$, 
where $\phi$ is the inflaton and $\chi$ the reheat field.
In this model the non-adiabatic pressure perturbation is given on large 
scales by
\begin{equation}
\delta p_{\rm nad} =
\frac {-m^2\phi\,\dot{\delta\chi}^2 + \ddot\phi
g^2\phi^2\delta\chi^2}{3H\dot\phi}
\,,
\end{equation}
where the long-wavelength solutions for vacuum fluctuations in the
$\phi$ field obey the adiabatic condition
$\delta\phi/\dot\phi=\dot{\delta\phi}/\ddot\phi$ (see also \cite{GWBM}).
We stress that we still only consider first-order perturbations
in the metric and total density and pressure, but these include
terms to second-order in $\delta\chi$.
This yields the power spectrum for the non-adiabatic part of the
curvature perturbation \cite{LLMW}
\begin{equation}
\label{Pzetanad}
{\cal P}_{\zeta_{\rm nad}} 
\simeq {2^{9/2}3 \over \pi^5\mu^2} \left({\Phi \over m_{\rm Pl}}\right)^2 
\left({H_{\rm end} \over m}\right)^4 g^4 q^{-1/4} \left({k\over k_{\rm
end}}\right)^3  I \,,
\end{equation}
where the integral $I$ can be approximated to first order
by $I \simeq 0.86 (m\Delta t)^{-3/2}\exp[4\mu_0 m\Delta t]$ 
(see \cite{LLMW} for details in notation).
It follows from Eq.~(\ref{Pzetanad}) that the 
scale dependence of ${\cal P}_{\zeta_{\rm nad}}$ is $k^3$.
Hence there is no change in the curvature perturbation in this model
due to preheating on scales relevant for large scale structure 
formation.
But it has been shown in \cite{Green} that the steep $k^3$-spectrum
in this model leads to 
an over-production of primordial black holes on smaller scales.

\end{document}